\documentclass[twocolumn]{aastex631}

\begin{document}

\title{High-energy gamma-ray and neutrino emissions from interacting supernovae based on radiation hydrodynamic simulations: a case of SN 2023ixf}

\author[0000-0003-2579-7266]{Shigeo S. Kimura}
\affiliation{Frontier Research Institute for Interdisciplinary Sciences, Tohoku University, Sendai 980-8578, Japan}
\affiliation{Astronomical Institute, Graduate School of Science, Tohoku University, Sendai 980-8578, Japan}

\author[0000-0003-1169-1954]{Takashi J. Moriya}
\affiliation{National Astronomical Observatory of Japan, National Institutes of Natural Sciences, 2-21-1 Osawa, Mitaka, Tokyo 181-8588, Japan}
\affiliation{Astronomical Science Program, Graduate Institute for Advanced Studies, SOKENDAI, 2-21-1 Osawa, Mitaka, Tokyo 181-8588, Japan}
\affiliation{School of Physics and Astronomy, Monash University, Clayton, VIC 3800, Australia}



\begin{abstract}
 Recent observations of core-collapse supernovae revealed that the existence of dense circumstellar matter (CSM) around their progenitors is ubiquitous. Interaction of supernova ejecta with such a dense CSM is a potential production sight of high-energy cosmic rays (CRs), gamma-rays, and neutrinos. We estimate the gamma-ray and neutrino signals from SN 2023ixf, a core-collapse supernova occurred in a nearby galaxy M101, which exhibits signatures of the interaction with the confined dense CSM.
Using radiation-hydrodynamic simulation model calibrated by the optical and ultraviolet observations of SN 2023ixf, we find that the CRs cannot be accelerated in the early phase because the sharp velocity jump at the shock disappears due to strong radiation pressure. Roughly 4 days after the explosion, the collisionless sub-shock is formed in the CSM, which enables the CR production and leads to gamma-ray and neutrino emissions. The shock sweeps up the entire dense CSM roughly 9 days after the explosion, which ceases the high-energy radiation. Based on this scenario, we calculate the gamma-ray and neutrino signals, which have a peak around 9 days after the explosion. We can constrain the cosmic-ray production efficiency to be less than 10\% by comparing our prediction to the Fermi-LAT upper limits. Future multi-messenger observations with an enlarged sample of nearby supernovae will provide a better constraint on the cosmic-ray production efficiency in the early phases of supernovae.
\end{abstract}



\section{Introduction} \label{sec:intro}

Supernovae remnants (SNRs) are believed to be the origin of Galactic cosmic rays (CRs) at GeV to PeV energies \citep[e.g.,][]{2012SSRv..173..369H}. This paradigm is partly confirmed by gamma-ray observations in GeV-TeV energies \citep{Ackermann:2013wqa,2012ApJ...746...82F}. These observations revealed that gamma-ray spectra of nearby SNRs detected in TeV energies have cutoff or break at 1-10 TeV energies \citep{Aha13a}, indicating that they do not currently contain higher energy CRs. This causes a debate whether SNRs are indeed the origin of PeV CRs or not.

Recently, TeV-PeV gamma-ray and neutrino diffuse emissions from the Galactic plane are detected by Tibet AS$\gamma$ \citep{2021PhRvL.126n1101A}, Large High Altitude Air Shower Observatory (LHAASO; \citealt{2023PhRvL.131o1001C}), IceCube \citep{2023Sci...380.1338I}, and possibly Baikal Gigaton Volume Detector (Baikal GVD; \citealt{2025ApJ...982...73A}), confirming that PeV CRs are Galactic origin. Several types of Galactic objects are proposed as PeV cosmic-ray sources, such as star-forming region and superbubble \citep{2020SSRv..216...42B}, stellar-mass black holes \citep{2020MNRAS.493.3212C,2021ApJ...915...31K}, and Sgr A* \citep{FMK17a}.

Core-collapse supernovae (SNe) that interact with a dense circumstellar matter (CSM) are also a candidate of PeV CR sources \citep[e.g.,][]{MTO14a,2018MNRAS.479.4470M}. Recent observations of Type~II SNe, which consist of 70\% of core-collapse SNe, indicate that they commonly show the signatures of confined dense CSM \citep{2018NatAs...2..808F,2023ApJ...952..119B}. 
Such a dense CSM can efficiently amplify the magnetic fields in the shock. This results in efficient particle acceleration to higher energies, achieving PeV energies in some parameter space \citep{2021ApJ...922....7I}.
These CRs accelerated in the dence CSM will efficiently interact with CSM, which leads to efficient hadron-induced gamma-ray and neutrino emissions \citep{Murase:2010cu,2018PhRvD..97h1301M,2019ApJ...874...80M,2017MNRAS.470.1881P,2019ApJ...872..157W}. Thus, SNe with interaction signatures are good targets of multi-messenger observations.

Last year, SN 2023ixf occurred in a nearby galaxy, M101 \citep{2023ApJ...955L...8H,2023ApJ...954L..42J,2023PASJ...75L..27Y,2023ApJ...956...46S,2023ApJ...953L..16H,2023ApJ...956L...5B,2023ApJ...954L..12T,2024Natur.627..759Z,2024ApJ...969..126Y,2024Natur.627..754L,2024arXiv240807874H,10.1093/mnras/stae2012}. This SN showed clear signatures of CSM interactions in the initial stage, but neither GeV gamma rays nor neutrino are observed \citep{2024A&A...686A.254M,2023ATel16043....1T}. Using these non detection, we can put constraint on cosmic-ray production efficiency in the CSM-interaction phase. 

In this paper, we discuss hadron-induced gamma-ray and neutrino emission from SN 2023ixf based on radiation hydrodynamic (RHD) simulation models. CSM parameters of SN 2023ixf are calibrated using RHD models.
We extract physical quantities relevant for CR production and gamma-ray/neutrino emissions from these simulations. We calculate time evolution of the gamma-ray and neutrino signals and sho that the constraint on CR production efficiency obtained by gamma-ray data is not in tension with that obtained by SNR modeling.

\section{RHD model for SN 2023ixf} \label{sec:rhd}

In this section, we discuss the dynamics of SNe interacting with dense CSM using RHD simulations. The CSM-ejecta interactions form the forward and reverse shocks. These shocks heats up the fluid, which produces strong radiations. These photons push the CSM material and modify the dynamical structure. To accurately treat this non-linear phenomena, we need frequency-dependent RHD simulations to predict the lightcurves of SNe powered by the CSM interaction.

We perform one-dimensional RHD simulations using the frequency dependent radiation transfer code, \texttt{STELLA} \citep{blinnikov1998,blinnikov2000,blinnikov2006}. This code solves the basic RHD equations, including hydrodynamic equations and radiative transfer equation taking the frequency domain into account. This enables us to predict the multi-wavelength lightcurves without assuming the thermal photon spectrum as well as the physical quantities in the ejecta-CSM system.

\begin{figure}[t!]
\plotone{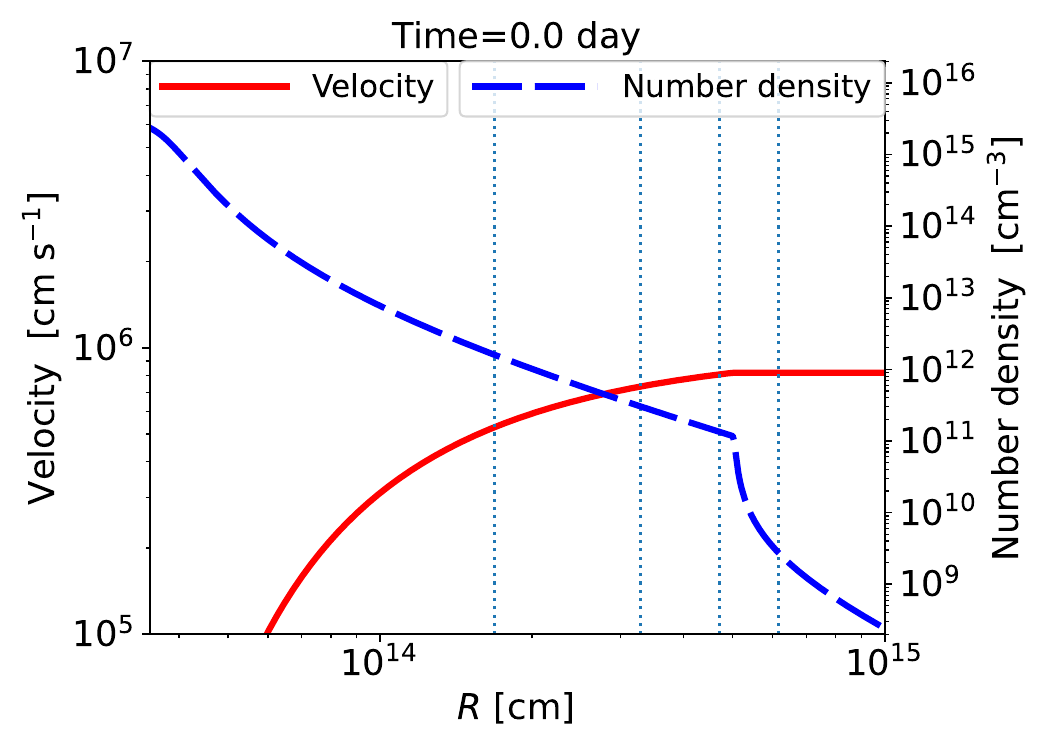}
\caption{The initial CSM velocity (red-solid) and number density (blue-dashed) profiles for the model that can explain the optical lightcurve of SN 2023ixf at various phases. The vertical lines indicate the shock radii for the panels in Fig. \ref{fig:hydro}.
\label{fig:initial}}
\end{figure}

We show the initial CSM density profile in Fig. \ref{fig:initial}.
We put a confined dense CSM component within a radius of $R<R_{\rm csm}=5\times10^{14}$ cm with the density profile following the $\beta$-law profile:$V_w=V_0+(V_\infty-V_0)(1-R_0/R)^\beta$, where $v_0$ and $V_w$ are the initial and terminal velocities of the wind, respectively, and $R_0$ cm is the wind launching radius \citep{2018MNRAS.476.2840M}.
We use the mass loss rate of $\dot{M}_w=0.01~M_\odot\rm~yr^{-1}$ with the wind velocity of $V_\infty=10^6\rm~cm~s^{-1}$, and $\beta=3$ \citep{2024arXiv240520989S}. $R_0$ is set to be the same with the stellar surface, and the value of $V_0$ does not affect our result as long as $V_0\ll V_\infty$ is satisfied.
The density structure settles to $n=\dot{M}_w/(4\pi m_p\mu R^2V_w)$, where $m_p$ is the proton mass and $\mu=1.26$ is the mean atomic weight. The mass of the confined CSM is $M_{\rm CSM}\simeq 0.7~M_\odot$ with this setup. The confined CSM is smoothly connected to a lower-density CSM whose average mass-loss rate is $\dot{M}_w=10^{-4}~M_\odot~\mathrm{yr^{-1}}$. The explosion of a red supergiant progenitor with the zero-age main sequence mass of 10~$M_\odot$ having the radius of 470~$R_\odot$ and the explosion energy of $2\times10^{51}$ erg can explain the optical lightcurves of SN 2023ixf \citep{2024arXiv240520989S,2024arXiv240600928M}.

\begin{figure*}[t!]
\plotone{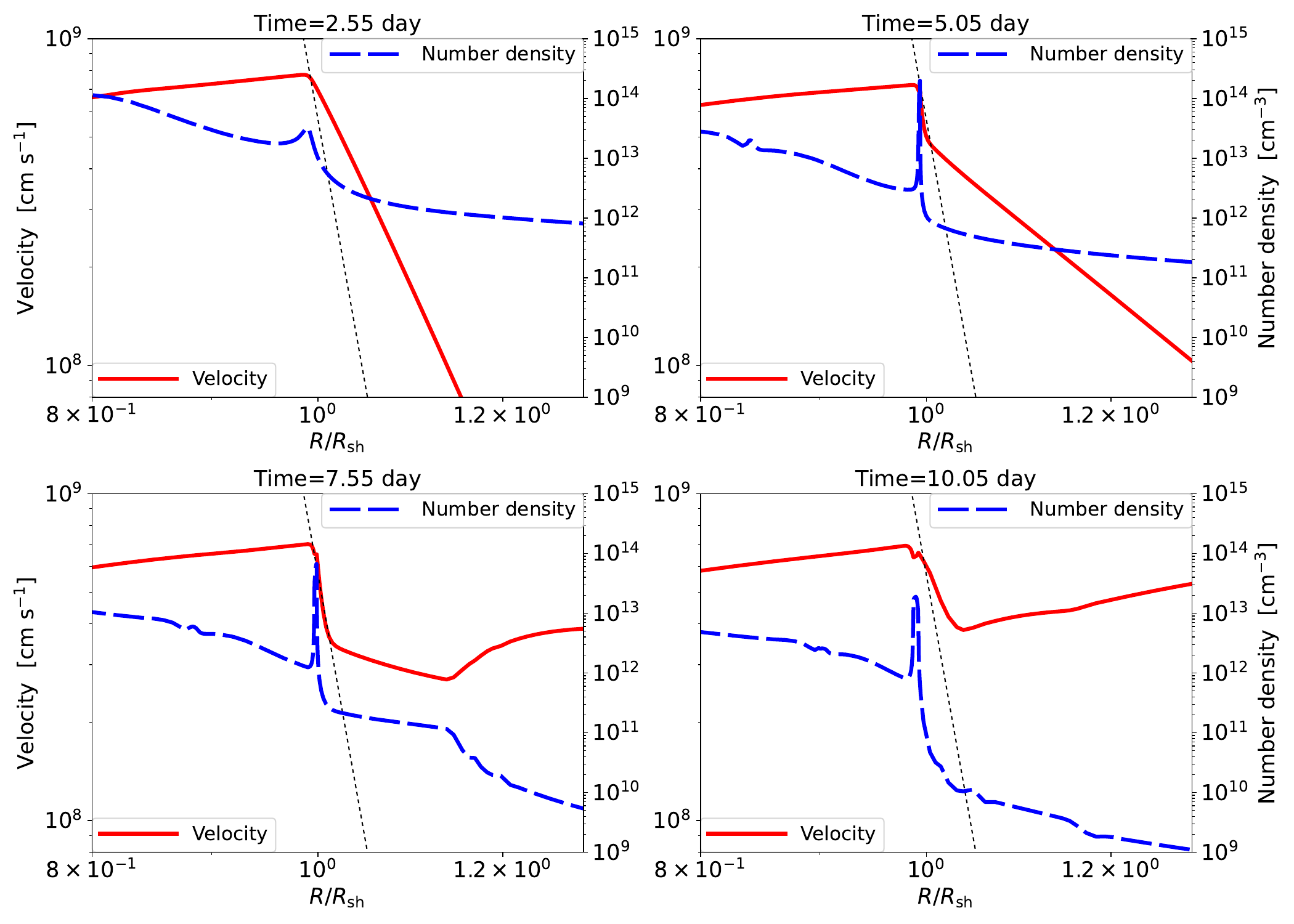}
\caption{The fluid Velocity (red-solid) and number density (blue-dotted) profiles around the shock for various times since explosion. The radius is normalized by the shock radius, and the thin-dotted lines ($\propto R^{-40}$) are shown to compare the sharpness of the shocks at different values of $t$.  For $t\lesssim4$ days, we cannot see the velocity jump because the strong radiation pressure pushes the unshocked CSM material. For $t\sim4-9$ days, we can see development of a collisionless sub-shock where cosmic rays can be efficiently accelerated. For $t\gtrsim9$ days, the collisinless shock disappears again because the mass of the unshocked CSM is so small that the radiation pressure can easily blow away it.
\label{fig:hydro}}
\end{figure*}

Fig. \ref{fig:hydro} shows the velocity and number density profiles around the shock at 2.55, 5.05 days, 7.55, 10.05 days after the core collapse.
At the initial stage of $t\lesssim4$ days, the shock released a large amount of energy due to high CSM density. This energy is efficiently converted to radiation energies. Then, the radiation pressure is strong enough to blow away the CSM near the shock, causing the gradual velocity change as seen in the panel for $t=2.55$ day. This situation is interpreted as the radiation mediated shock. Since there is no sharp velocity jump, CRs are unlikely to be accelerated in this phase.

At $t\sim4$ days, the density slightly decreases and radiations can partly escape from the system, which decreases the radiation pressure. In this situation, radiation pressure cannot push away all the CSM, forming a sub-shock mediated by plasma instability where a sharp velocity jump exists (see the panel for $t=5.05$ day in Fig. \ref{fig:hydro}). Thus, cosmic rays start to be accelerated around this time. The sub-shock grows in time, meaning that the amount of energy released at the collisionless sub-shock increases with time. One can see that the density at the shocked CSM is 2-3 orders of magnitude higher than that in the unshocked CSM for $t\gtrsim4$ days, indicating the formation of radiative shock where the production and escape of the photons are efficient.

For $t\sim6-9$ days, the CSM density becomes low enough for radiations to efficiently escape from the system. Then, the most of the available shock energy is released at the collisionless shock as seen in the panel for $t=7.55$ day in Fig. \ref{fig:hydro}. Even in this stage, the density is still high enough to produce a large amount of cosmic rays and efficiently produce gamma-rays and neutrinos via hadronuclear interactions as seen in the following sections.

At $t\sim9$ days, the shock has completely swept up the confined dense CSM component. For $t\gtrsim9$ days, the shock is located at the lower density CSM component. Just before the shock arriving there, a large amount of photons pass through the low-density CSM, which blows away it. Then, the sharp velocity jump disappears again (see the panel of $t=10.05$ day in Fig. \ref{fig:hydro}), which causes to cease CR acceleration. Thus, we do not expect gamma-ray and neutrino signals after $t\gtrsim9$ days.

\section{Particle Acceleration at shocks} \label{sec:accel}

In this section, we discuss CR acceleration in the collisionless sub-shock by estimating the acceleration and loss timescales. 
Since the sharp collisionless shock appears only for $t\sim4-9$ days, we hereafter focus on this time window. 
Based on the diffusive shock acceleration theory, the particle acceleration timescale at the shock depends on the fluid velocity at the shock rest frame \citep{Dru83a}. We denote the velocity of the unshocked fluid at the shock rest frame as $V_{\rm sh}$, because this corresponds to the shock velocity for typical supernova remnants. Assuming a strong shock, we can write $V_{\rm sh}=(4/3)\Delta V$, where $\Delta V$ is the jump of the fluid velocity between the shocked and unshocked media.  Our RHD simulation is performed in the observer frame where the unshocked fluid velocity is non-zero. Thus, $V_{\rm sh}$ does not match the time derivative of the shock position in the observer frame.

\begin{figure}[t!]
\includegraphics[width=1.0\columnwidth]{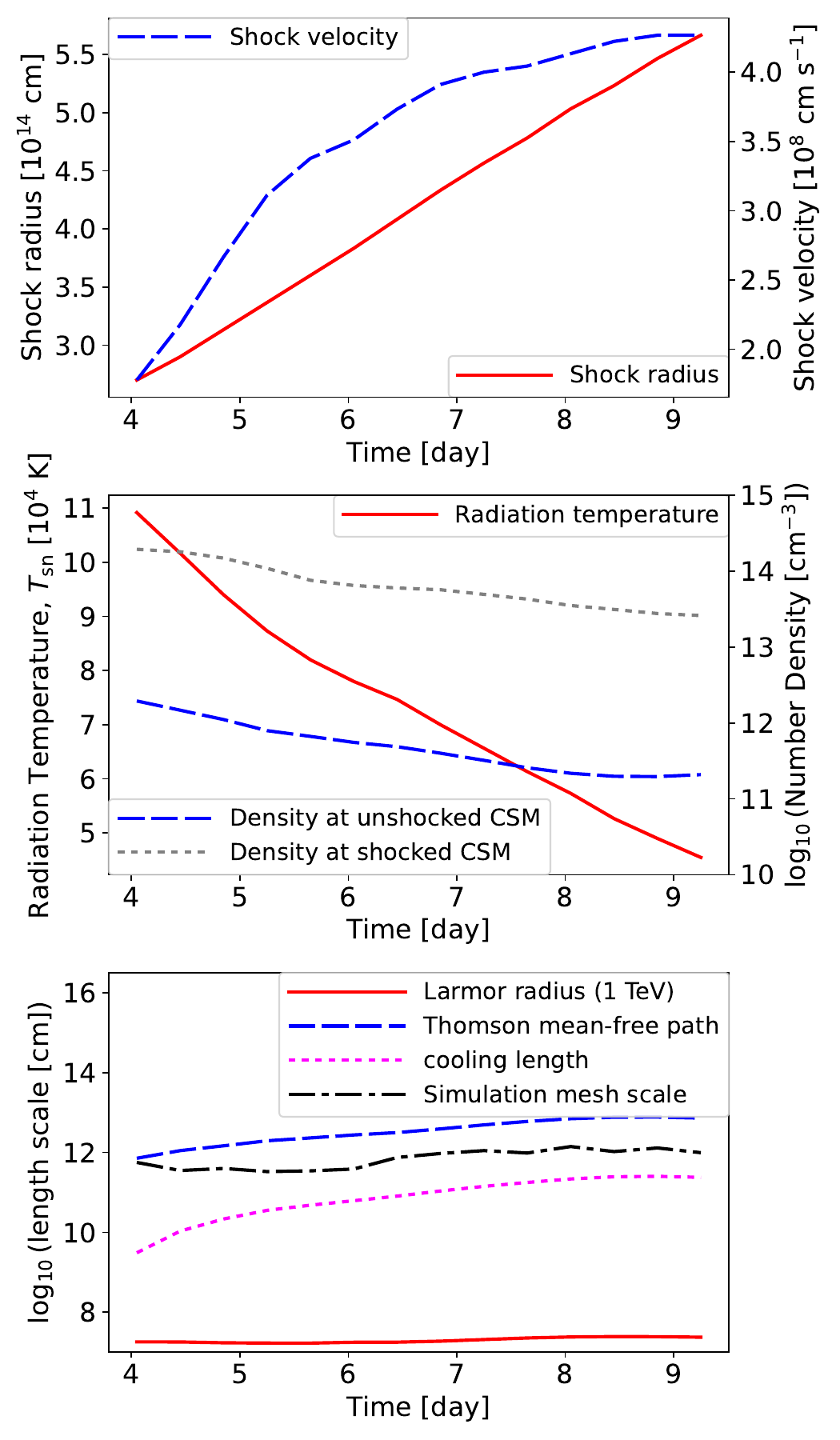}
\caption{Time evolutions of physical quantities around the shock extracted from the RHD simulation for SN 2023ixf. The top panel shows the shock radius (red-solid) and the velocity of unshocked fluid at the shcok-rest frame ($V_{\rm sh}$; blue-dashed). The middle panel shows the radiation temperature at the shocked region (red-solid) and the density at the shocked (grey-dotted) and unshocked (blue-dashed) CSM. The bottom panel shows the Larmor radius for 1-TeV protons (red-solid), Thomson mean-free path (blue-dashed), cooling length (magenta-dotted), and grid scale for the RHD simulation (black-dotted-dashed) at the shocked region. 
\label{fig:shock}}
\end{figure}

Fig. \ref{fig:shock} shows the time evolution of physical quantities at the shock. As seen in the top and middle panels, the physical quantities do not change much; The shock radius is $R_{\rm sh}\sim3\times10^{14}-6\times10^{14}$ cm, the shock velocity $V_{\rm sh}\sim2\times10^8-4\times10^8\rm~cm~s^{-1}$, the radiation temperature at the shocked region $T_{\rm sn}\sim5\times10^4-1\times10^5$ K, and the densities at the shocked and un-shocked regions are $n_{\rm down}\sim2\times10^{13}-1\times10^{14}\rm~cm^{-3}$ and $n_{\rm up}\sim2\times10^{11}-1\times10^{12}\rm~cm^{-3}$, respectively.

In order to estimate timescales relevant for CR production, we need to estimate the physical quantities of the CR acceleration site.
The width of the shock, i.e., the region where a sharp velocity jump occurs, should be of the order of the plasma skin depth or the gyration radius for thermal particles, which is the shortest length scale relevant for CR acceleration. The physical quantities in the immediate shock downstream should be described by the adiabatic shock jump condition.
The shocked fluid will cool by free-free emission in a cooling timescale, which leads to a strong compression in a cooling length of the fluid. We can see the compression in our RHD simulation (see Fig. 2). On the other hand, the CR acceleration should take place within the mean-free path of the CR particles from the shock, which are evaluated using the gyration radius of CR particles. Therefore, we need to compare the cooling length of the fluid and the gyration scale of the CR particles to judge whether CR acceleration should take place in the adiabatic or radiative shock.

We evaluate several length scales around the shock to discuss whether the CR acceleration and loss take place in the adiabatic or radiative shock. We estimate the Thomson mean-free path by $l_{\rm Th}\approx 1/(4\sigma_Tn_{\rm up})$, where $\sigma_T$ is the Thomson crosssection. The cooling length is estimated to be $\l_{\rm ff}\approx n_{\rm up}k_BT_{\rm ad,sh}V_{\rm sh}/\Lambda_{\rm ff}$, where $k_BT_{\rm ad,sh}=(3/16)\mu m_pV_{\rm sh}^2$ is the temperature at the shock downstream and $\Lambda_{\rm ff}\approx 1.7\times10^{-27}T_{\rm ad,sh}(4n_{\rm up})^2 \rm~erg~s^{-1}~cm^{-3}$ is the cooling rate by free-free emission \citep[e.g.,][]{rl79}. The Larmor radius is given by $r_L=E_p/(eB)$, where $E_p$ is the proton energy, $e$ is the elementary charge, $B\approx\sqrt{32\pi\epsilon_B\mu m_pn_{\rm up}V_{\rm sh}^2}\simeq82{\rm~G}~n_{\rm up,11.5}^{1/2}V_{\rm sh,8.5}\epsilon_{B,-3}^{1/2}$ is the magnetic field at the immediate downstream, and $\epsilon_B$ is the magnetic-field amplification parameter. Hereafter, we use the notation of $Q_x=Q/10^x$ in cgs unit.

The bottom panel of Fig. \ref{fig:shock} shows the comparison of these length scales together with the simulation mesh scale. We find that the Larmor radius is  smaller than the cooling length by a few orders of magnitude even at TeV energies, meaning that the particle acceleration should occur at the adiabatic shock. The cooling length is the second smallest, and the simulation mesh scale follows. The Thomson mean-free path is the longest of the four. This means that the radiation mediated shock should be resolved by several mesh, ensuring that the velocity structure around the shock seen in Fig. \ref{fig:hydro} is physical\footnote{We should note that the mesh scale around the shock is longer than the Thomson mean-free path at the very early phase of $t\lesssim1$ day. In this phase, the simulation result exhibits a sharp velocity jump at the shock. However, this is due to the lack of the resolution, and the shock is mediated by radiation, rather than the plasma instability. Thus, we do not expect CR acceleration for $t\lesssim1$ day.}.

\begin{figure}[t!]
\plotone{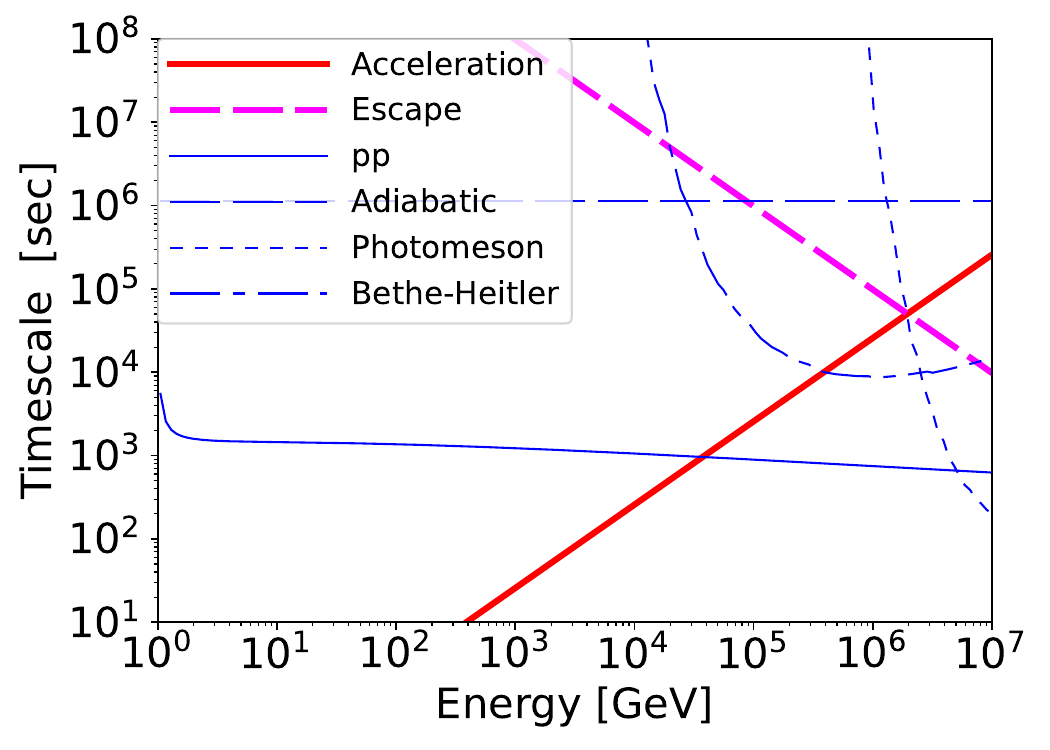}
\caption{Comparison of acceleration, escape, and cooling timescales as a function of energies for $t=7.25$ days.
\label{fig:timescale}}
\end{figure}

Next, we estimate the acceleration, escape, and cooling timescales for CR protons to evaluate the maximum energy. The acceleration and escape timescales are estimated to be $t_{\rm acc}\approx (20/3)(\eta r_L/c)(V_{\rm sh}/c)^{-2}$ \citep{Dru83a} and $t_{\rm esc}=D_{\rm sh}^2/(\eta r_L c/3)$, where $\eta$ is the Bohm factor, $c$ is the speed of light, and $D_{\rm sh}$ is the thickness of the shocked region. A lower value of $\eta\sim1-10$ is implied by X-ray observations of SNRs \citep{2005ApJ...621..793B,2008ARA&A..46...89R,2020ApJ...904..188K}, and we fix $\eta=1$ throughout this study for simplicity. We approximate $D_{\rm sh}\approx0.01R_{\rm sh}$, which is consistent with the RHD simulation result\footnote{The thickness of the shocked region becomes thinner and thinner in the RHD simulation because of the efficient radiative cooling. We avoid this by introducing a smearing term \citep{blinnikov1998,2013MNRAS.428.1020M}. In reality, the magnetic pressure and multi-dimensional motion will play important roles to determine $D_{\rm sh}$, which demand magnetohydrodynamic (MHD) as well as multi-dimensional simulations.}.

We consider hadronuclear interaction ($pp$; $p+p\rightarrow p+p+\pi$), photomeson production ($p\gamma$; $p+\gamma\rightarrow p+\pi$), Bethe-Heitler process (BH; $p+\gamma\rightarrow p+e^++e^-$), and adiabatic expansion as cooling processes.
$pp$ and $p\gamma$ processes produce neutral and charged pions. Neutral pions decay to two gamma-rays ($\pi^0\rightarrow 2\gamma$), and charged pions decay to neutrinos and electrons/positrons ($\pi^\pm \rightarrow 3\nu + e^\pm$). Thus, if CRs are accelerated at the dense CSM, we expect efficient gamma-ray and neutrino production.

The $pp$ and adiabatic cooling rates are estimated to be $t_{pp}^{-1}\approx 4n_{\rm up}\sigma_{pp}\kappa_{pp}c$ and $t_{\rm ad}^{-1}\approx V_{\rm sh}/R_{\rm sh}$, respectively, where $\sigma_{pp}$ and $\kappa_{pp}$ are the crosssection and inelasticity for hadronuclear interactions. We use the density at the immediate downstream of the shock, $4n_{\rm up}$, when estimating $t_{pp}$. We use $\sigma_{pp}$ given in \citet{2014PhRvD..90l3014K} and $\kappa_{pp}\approx 0.5$. The photomeson and Bethe-Heitler cooling rates, $t_{\rm p\gamma}$ and $t_{\rm BH}$, are estimated by the same method with \cite{2019PhRvD.100h3014K}, where we use the crosssection for photomeson production by \cite{MN06b}  and Bethe-Heitler process are given in \cite{SG83a,CZS92a}. The photon fields are assumed to be Planck distribution with the temperature $T_{\rm sn}$ obtained by the RHD simulation. This approximates the photon fields around the shock accurate enough for our purpose. 

Fig. \ref{fig:timescale} plots these timescales as a function of CR energies. We see that the hadronuclear interaction is the most efficient loss process and limit the proton acceleration. The resulting maximum energy is written as
\begin{equation}
E_{p,\rm max}\approx\frac{3eBV_{\rm sh}^2}{80\eta n_{\rm up}\sigma_{pp}\kappa_{pp}c^2}\simeq22 n_{\rm up,11.5}^{-1/2}V_{\rm sh,8.5}^3\epsilon_{B,-3}^{1/2}\rm~TeV. \label{eq:Emax}
\end{equation}
Thus, SN 2023ixf cannot be a PeVatron because of the efficient cooling by hadronuclear interactions. The maximum energy is higher at the later phase because the shock is faster and density is lower (see Fig. \ref{fig:shock}). To achieve PeV energies, we need to consider a a lower density CSM, i.e., a lower $\dot{M}_w$ or a larger $R_{\rm csm}$.
Fig. \ref{fig:timescale} also indicates the high pion production efficiency, $f_{pp}=t_{\rm cool}/t_{pp}\simeq1$, meaning that all the CR energies are converted to the secondary particles, mainly gamma-rays and neutrinos.

\section{High-energy emissions from SN 2023ixf} \label{sec:radiation}

In this section, we discuss high-energy gamma-ray and neutrino signals expected from SN 2023ixf. To calculate the neutrino and gamma-ray spectra, we need to obtain the number spectrum of CR protons, $N_{E_p}$. Since the acceleration and cooling timescales are much shorter than the dynamical timescale (see Fig. \ref{fig:timescale}), we assume that the proton spectrum reaches a steady state. Then, the transport equation for CR protons are written as
\begin{equation}
 \frac{\partial}{\partial E_p}\left(\frac{E_pN_{E_p}}{t_{\rm cool}}\right)+\dot{N}_{p,\rm inj}-\frac{N_{E_p}}{t_{\rm esc}}=0,\label{eq:transport}
\end{equation}
where $t_{\rm cool}^{-1}=t_{pp}^{-1}+t_{\rm ad}^{-1}+t_{p\gamma}^{-1}+t_{\rm BH}^{-1}$ is the total cooling rate and $\dot{N}_{p,\rm inj}$ is the injection term. Considering the diffusive shock acceleration by a strong shock, we use the power-law CR spectrum with an exponential cutoff: $\dot{N}_{p,\rm inj}=\dot{N}_0(E_p/E_{p,\rm max})^{-s_{\rm inj}}\exp(-E_p/E_{p,\rm max})$, where $\dot{N}$ is the normalization factor and $s_{\rm inj}$ is the injection spectral index. The normalization factor, $\dot{N}_0$ is determined by 
\begin{eqnarray}
L_p=\int E_p\dot{N}_{p,\rm inj}dE_p = 4\pi \epsilon_pR_{\rm sh}^2 \mu m_p n_{\rm up} V_{\rm sh}^3  \label{eq:Lp}\\
\simeq 2.6\times10^{42}\epsilon_{p,-1}R_{\rm sh,14.5}^2n_{\rm up,11.5}V_{\rm sh,8.5}^3\rm~erg~s^{-1}\nonumber
\end{eqnarray}
where $\epsilon_p$ is the CR production efficiency. The resulting gamma-ray and neutrino flux is proportional to $\epsilon_p$, and we set $\epsilon_p=0.1$ as a reference value.
This value is widely used because $\sim10\%$ of SN kinetic energy needs to be converted to CRs in order to explain the Galactic cosmic-ray observations \citep[e.g.,][]{Hil05a}. 
We fix $s_{\rm inj}=2$ throughout this paper for simplicity, but the spectral index does not have strong influence on the GeV gamma-ray flux if we take the electromagnetic cascade into account \citep{2019ApJ...874...80M}. The neutrino flux at the IceCube band would be lowered if we use a softer spectral index suggested by radio observations of supernovae \citep{2012ApJ...758...81M} and PIC simulations \citep{2020ApJ...905....2C}. 

We numerically solve Eq. (\ref{eq:transport}) and obtain $N_{E_p}$. Then, we calculate the gamma-ray and neutrino spectrum using the method given by \cite{kab06}. Since the neutrino and gamma-ray production is dominated by hadronuclear interactions, we can neglect the contributions by the other processes. We take the neutrino oscillation into account using Eq. (47) in \cite{bec08}.

The neutrinos are freely escape from the system. On the other hand, the gamma-rays are attenuated by Bethe-Heitler process with the CSM and Breit-Wheeler processes ($\gamma\gamma$; $\gamma+\gamma\rightarrow e^++e^-$) with the thermal photons of temperature $T_{\rm sn}$. We estimate the optical depth for these process,  $\tau_{\rm BH}\approx n_{\rm up}\sigma_{\rm BH}R_{\rm sh}$ and $\tau_{\gamma\gamma}\approx \int N_{E_\gamma}\sigma_{\gamma\gamma}R_{\rm sh}dE_\gamma$, where $N_{E_\gamma}$ is the number spectrum of photons, $E_\gamma$ is the photon energy, and $\sigma_{\gamma\gamma}$ is the crosssection of Breit-Wheeler process. We use the $\sigma_{\gamma\gamma}$ given in \cite{cb90} and suppresses the gamma-ray flux using the suppression factor:
\begin{equation}
 F_{\rm att}=\frac{1-\exp(-\tau_{\gamma\gamma})}{\tau_{\gamma\gamma}}\exp(-\tau_{\rm BH}).
\end{equation}
This suppression factor strongly depends on the gamma-ray energy, $E_\gamma$. Our treatment of Breit-Wheeler attenuation ignores the geometrical effect, which might affect the shape of the high-energy tail in gamma-ray spectra. 

The thermal photon with $T_{\rm sn}\sim3\times10^4$ K interact with the gamma-ray photons of
\begin{equation}
 E_{\gamma\gamma,\rm sn}\approx \frac{4m_e^2c^4}{2.8k_BT_{\rm sn}}\simeq 87 T_{\rm sn,0.7}^{1/2}R_{\rm sh,14.5}^{-1/2} \rm~GeV.\label{eq:Egam}
\end{equation}
This energy is lower than the higher cutoff energies of pion-decay gamma-rays, $E_{\gamma,\rm max}\sim0.1E_{p,\rm max}\sim2\rm~TeV$ (see Eq. \ref{eq:Emax}), and  thus, electron-positron pairs are efficiently produced. These electron-positron pairs emit gamma-rays via inverse Compton scattering, which initiate the electromagnetic cascades. We approximately calculate the electromagnetic cascades by the method in \cite{2020ApJ...905..178K}, and resulting photon fields are calculated iteratively. The photons produced by the cascades make $E_{\gamma\gamma,\rm sn}$ lower than that estimated by Eq. (\ref{eq:Egam}).

\begin{figure}[t!]
\plotone{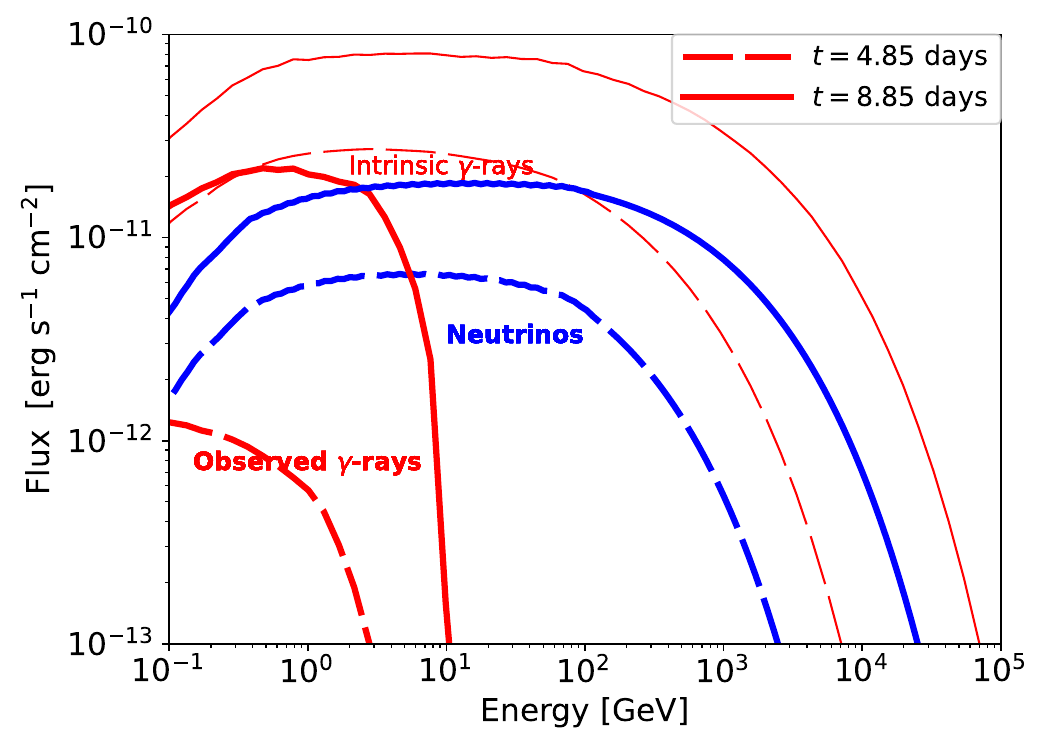}
\plotone{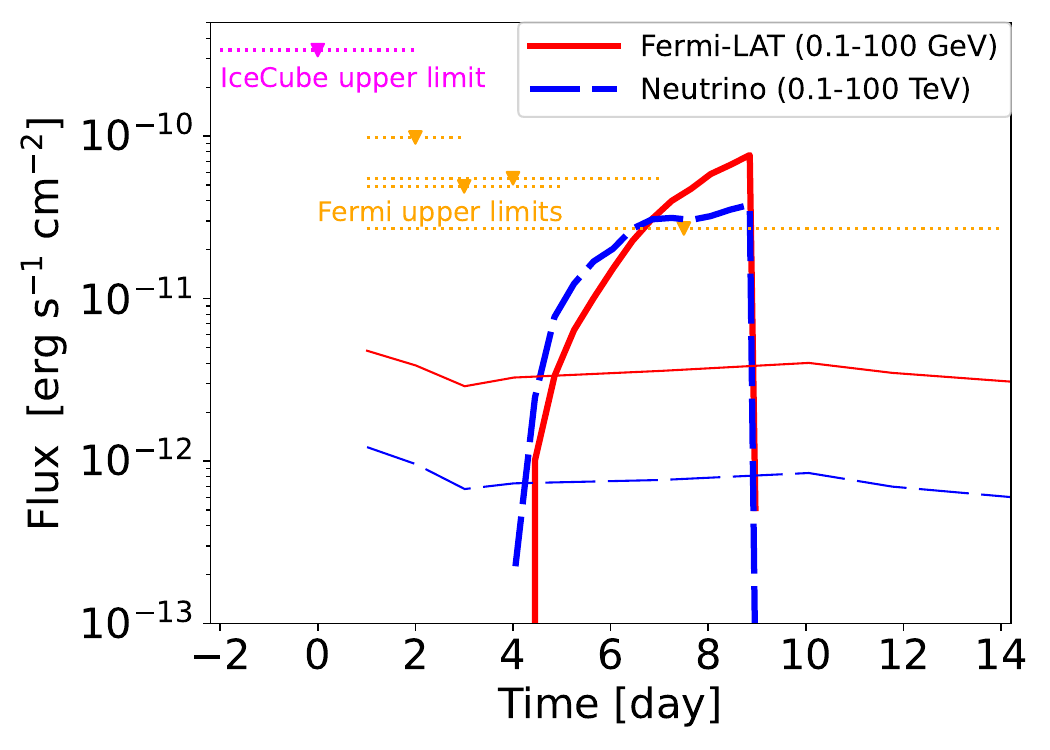}
\caption{Top: Intrinsic gamma-ray (thin-red), observed gamma-ray (thick-red), and muon-neutrino (thick-blue) spectra for our model. The solid and dashed lines show the spectra for $t=4.85$ and 8.85 days, respectively. The gamm-rays are strongly attenuated because of Breit-Wheeler and Bethe-Heitler processes.
Bottom: Gamma-ray (red) and neutrino (blue) lightcurves in Fermi-LAT and IceCube energy bands for our model. The thick and thin curves are emissions by optical and X-ray motivated CSMs, respectively. The inverted triangles with horizontal dotted lines show the upper limits obtained by Fermi-LAT (orange) and IceCube (magenta). 
\label{fig:lc}}
\end{figure}

The upper panel of Fig. \ref{fig:lc} shows the gamma-ray and neutrino spectra at $t=4.85$ and 8.85 days. The gamma-rays are significantly attenuated due to Bethe-Heitler and Brie-Wheeler processes below and above $\sim 10$ GeV, respectively. The attenuation by Bethe-Heitler process at GeV energies is severe in the early phase, but it becomes marginal in the later phase. The attenuation by Breit-Wheeler process is always severe above 10 GeV, and it was very challenging to detect SN 2023ixf with air-cherenkov detectors, even with Cherenkov Telescope Array. The neutrino flux has cuohtoff around 0.1-1 TeV, depending on the phase, which is lower than the sensitive energy range of IceCube. 

The bottom panel shows the lightcurves of gamma-rays in the Fermi band and neutrino in the IceCube band. We consider that after $t>9.25$ days, CR acceleration is ceased, and gamma-ray and neutrino lightcurves experience an exponential decay with the decay timescale equal to $t_{pp}$. We also plot the upper limit obtained by Fermi-LAT \citep{2024A&A...686A.254M} and IceCube \citep{2023ATel16043....1T}. Our model prediction with the reference value of $\epsilon_p=0.1$ is comparable to the Fermi-LAT upper limit with a longer time-window analyses. On the other hand, the predicted neutrino flux is much lower than the current neutrino upper limit. Even with near-future detectors, it is challenging to detect high-energy neutrino signals from nearby SN 2023ixf-like objects. This conclusion is consistent with previous estimate \citep{2023ApJ...956L...8K}.

\begin{figure}[t!]
\plotone{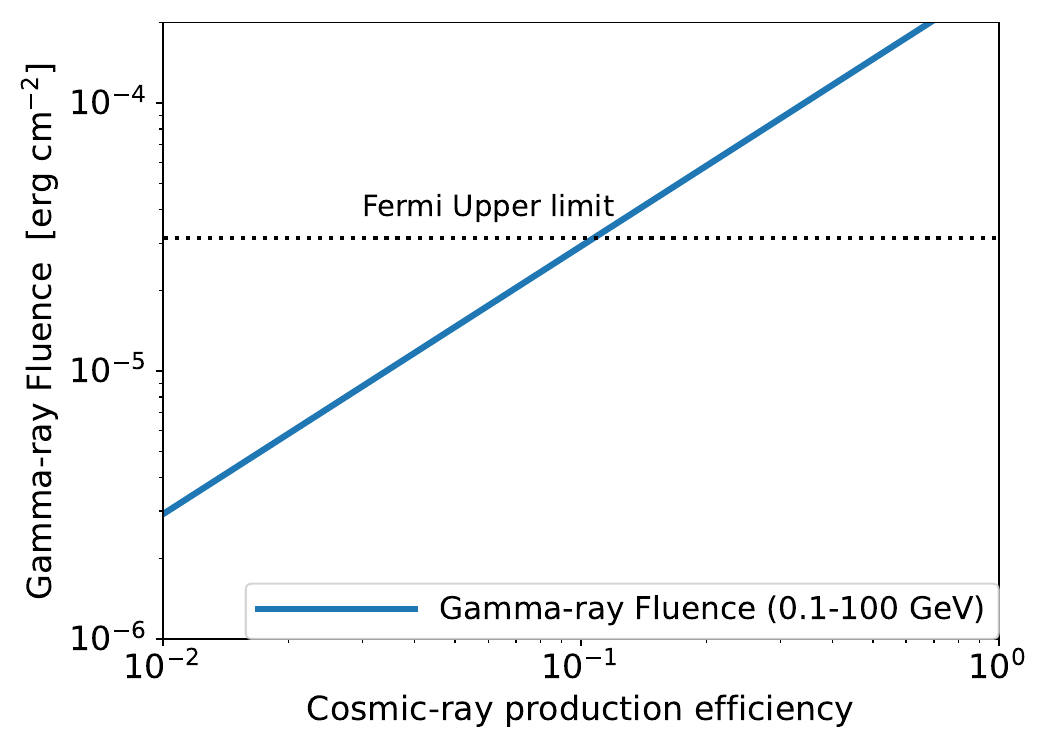}
\caption{The predicted gamma-ray fluence in the Fermi-LAT band as a function of cosmic-ray proton production efficiency, $\epsilon_p$. We can see that the canonical value of $\epsilon_p=0.1$ is still allowed in our modeling. 
\label{fig:epsp}}
\end{figure}

Since our model is calibrated using the RHD simulation with the optical data, we only have a single free parameter, $\epsilon_p$\footnote{We have two other parameters, $\epsilon_B$ and $\eta$, but they do not affect the gamma-ray flux at the Fermi band. They will affect the maximum energy of the protons, so $\epsilon_B$ and $\eta$ would be relevant when we discuss the IceCube limit.}. Fig. \ref{fig:epsp} shows the predicted gamma-ray fluences as a function of $\epsilon_p$, and compare it with the 14-day upper limit by Fermi. We can see that the predicted fluence overshoots the upper limit if we consider $\epsilon_p\gtrsim0.1$. Thus, the canonical value of $\epsilon_p\sim0.03-0.1$ is marginally allowed in diffusive shock acceleration. Analyses with an adequate time window might improve the constraint.

\section{Discussion}

SN 2023ixf is detected in X-rays \citep{2023ApJ...952L...3G,2024ApJ...963L...4C}.
Our RHD model calibrated with the optical data cannot reproduce the X-ray data because the dense CSM completely attenuates the X-rays emitted from the shocked fluid.
To reproduce the X-ray data, we need a lower-density CSM. The RHD model calibrated by the X-ray observation is $n_{\rm up}\sim 1.4\times10^{10}\rm ~cm^{-3}$ at $R\sim10^{14}$ cm, which is much lower than that estimated by the optical data given in Fig. \ref{fig:hydro}.
The CSM around SN~2023ixf is found to be aspherical \citep{2024arXiv240520989S} and both CSM components are likely to exist together. Here, we discuss the gamma-ray and neutrino emission from the interaction with the CSM estimated by X-ray observations.

With the X-ray motivated CSM, the shock should become collisionless at $R_{\rm sh}\sim10^{14}$ cm. At this point, the CR luminosity is estimated to be $L_p\simeq 3.8\times10^{41} \epsilon_{p,-1}R_{\rm sh,14}^2n_{\rm up,10}V_{\rm sh,9}^3 \rm~erg~s^{-1}$. This value is about an order of magnitude smaller than that given in Eq. (\ref{eq:Lp}), and thus, the gamma-ray and neutrino emissions from the X-ray motivated CSM are much fainter than those by optical motivated ones. We plot the gamma-ray and neutrino lightcurves for models with X-ray motivated CSM in the bottom panel of Fig. \ref{fig:lc}, where we see that these components are sub-dominant compared to the optical motivated ones.  Nevertheless, these emissions could be relevant signals because this component emit gamma-rays and neutrinos for longer timescales. Based on the RHD simulation model, we confirm that the gamma-ray and neutrino emissions do not decay until $t\sim14$ days. The systematic investigation on the gamma-ray and neutrino detectability with various RHD simulation models are beyond the scope of this paper.

\cite{2024A&A...686A.254M} reported that the CR production efficiency in SN 2023ixf should be less than 1\% using the Fermi-LAT upper limit. They consider CR production even when the collisionless shock does not exist, i.e., putting a constraint on $\epsilon_p$ using the full time window. On the other hand, our model  takes into account the CR production only when collisionless shocks are developed, i.e., putting a constraint on $\epsilon_p$ using the adequate time window. This difference leads to constraints on $\epsilon_p$ that differ by an order of magnitude.
We need to determine the time window of CR acceleration through RHD modelings to understand the physics of particle acceleration. 

High-energy neutrino and gamma-ray emissions from SN 2023ixf are discussed in the previous literature. \cite{2023ApJ...955L...9G} adopt the choked jet scenario where they consider jets below the photosphere as a neutrino production site \citep[e.g.,][]{HKN18a}. This is a different scenario from our CSM interaction scenario. 
\cite{2024PhRvD.109j3020M} discusses gamma-ray and neutrino emission from SN 2023ixf using a similar CSM interaction scenario, but his calculation assumes a lower density CSM. The CSM density is parameterized using $D_*=\dot{M}/(4\pi V_w)$, and $D_*=0.1$ and 0.003 are used, which leads to $\epsilon_p\lesssim0.2$. Our model corresponds to $D_*\simeq 150$. This difference causes qualitatively different time evolution in neutrino and gamma-ray signals.
\cite{2024JCAP...04..083S} also discuss the gamma-ray and neutrino production from SN 2023ixf based on the CSM interaction scenario with a parameter set similar to that in our model. However, the radiation-mediated condition and electromagnetic cascades are ignored in his calculation. This causes strong gamma-ray and neutrino signals just 1 day after the explosion, which should not occur if we take into account radiation-mediated condition. Also, ignoring the electromagnetic cascade would cause the weaker attenuation by Breit-Wheeler process.
In addition, \cite{2025arXiv250303699C} discusses neutrino emission from SN 2023ixf by constructing semi-analytic model. Their model uses a lower density and extended CSM compared to ours, which leads to lower neutrino luminosity, although their conclusions for neutrino detectability is roughly consistent with ours. They discuss neutrino emission from the beginning to the late phase without discussing gamma-ray signals. On the other hand, we discuss both neutrino and gamma-ray emissions focusing on the early phase at which these fluxes are higher.

We ignored the back reactions of CRs to the shock dynamics. Since our results indicate that the CR carries only less than 10\% of the shock energy, the effect of CRs on hydrodynamic quantities are unlikely to be strong and observable \citep{2022PASJ...74.1022S}. On the other hand, CRs can induce a electric current when they diffuse out from the shock, which will lead to enhancement of magnetic fields \citep{Bel04a}. Also, CRs would produce a precursor and post-cursor around the shock with a scale of their gyration radii, which would affect the index of the CR spectrum \citep{2020ApJ...905....2C}. These effect could be taken into account by modeling the interacting SNe with radiation MHD-PIC simulations \citep[see, e.g.,][for MHD-PIC method]{BCS15a}, which is left as a future work. 

\section{summary}

We investigate the gamma-ray and neutrino production in a nearby supernova, SN 2023ixf, using a RHD simulation model calibrated by the optical lightcurve data of SN 2023ixf \citep{2024arXiv240520989S}. By extracting the physical quantities around the shock using the simulation, we find that the CR acceleration is efficient only for 4-9 days after the explosion. We numerically calculate the gamma-ray and neutrino signals during the time window and find that our model prediction is marginally consistent with the Fermi-LAT and IceCube upper limit with a canonical value of CR production efficiency, $\epsilon_p=0.1$. Our model does not have free parameters that can significantly change the gamma-ray flux except for $\epsilon_p$, and our model constrain the CR production efficiency to be $\epsilon_p\lesssim0.1$. Future observations with an enlarged sample of nearby SNe will be able to determine or constrain the CR efficiency near future.

 \begin{acknowledgments}
  We thank Kohta Murase and Ryo Yamazaki for useful discussions.  
  This work is partly supported by JSPS KAKENHI grant Nos. 22K14028, 21H04487, 23H04899 (S.S.K.), 24K00682, 24H01824, 21H04997, 24H00002, 24H00027, 24K00668 (T.J.M.) and the Tohoku Initiative for Fostering Global Researchers for Interdisciplinary Sciences (TI-FRIS) of MEXTs Strategic Professional Development Program for Young Researchers (S.S.K).
  TJM is supported by the Australian Research Council (ARC) through the ARC's Discovery Projects funding scheme (project DP240101786). Numerical computations were in part carried out on PC cluster at the Center for Computational Astrophysics, National Astronomical Observatory of Japan.
 \end{acknowledgments}

%

\vspace{5mm}








\bibliography{sample631,moriya}{}
\bibliographystyle{aasjournal}



\end{document}